\documentstyle[epsfig,graphicx,11pt,epsf]{article}

\newcommand{\nn}{\nonumber}

\newcommand{\be}{\begin{equation}}
\newcommand{\ee}{\end{equation}}
\newcommand{\ba}{\begin{eqnarray}}
\newcommand{\ea}{\end{eqnarray}}

\begin{document}
\begin{center}
{\Large{\bf The $\phi NN$ coupling from chiral loops
  }}

\vspace{0.3cm}

\end{center}

\vspace{1cm}

\begin{center}
{\large{J. E. Palomar and E. Oset}}
\end{center}

\begin{center}
{\small{\it Departamento de F\'{\i}sica Te\'orica and IFIC, \\
Centro Mixto Universidad de Valencia-CSIC, \\
Ap. Correos 22085, E-46071 Valencia, Spain}}

\end{center}

\vspace{1cm}

\begin{abstract}
Starting from effective Lagrangians which combine a gauge formulation of Vector
Meson Dominance with Chiral Lagrangians, the coupling of the $\phi$ to the
nucleon, which is zero at tree level due to the OZI rule, is calculated 
perturbatively considering loop contributions to the electric and magnetic form
factors. We obtain reasonably smaller values for both form factors 
than those for $\rho NN$ and consistent with the expected 
order of magnitude of the OZI rule violation. The role of the $\phi -\omega$
mixing is also investigated.
\end{abstract}

\section{Introduction}
  A general formulation of vector meson couplings to pseudoscalar 
  mesons and baryons can be constructed combining elements of Vector Meson
Dominance and SU(3) chiral Lagrangians \cite{Kaymakcalan:1984bz,Borasoy:1995ds,Klingl:1996by,Klingl:1997kf}, hence
placing the $\phi$ and the $\rho$ on the same footing. Yet,  the
$\rho$ and $\phi$ couple to the nucleon in a very different way, since the
Lagrangians are consistent with the OZI rule and thus the $\phi$, which stands
for a $s\bar{s}$ state in this formulation, does not couple to the nucleon nor
to pions at the tree level. The same Lagrangians, however, allow one to perform
perturbative calculations to account for loop contributions to the $\phi$
couplings, involving kaons and hyperons to which the $\phi$ couples naturally.
 One, nevertheless, still expects the couplings to be small since the OZI rule 
 should not be 
much violated. One of the reactions where the OZI rule shows up, drastically
reducing the decay rate, is the $\phi \to \pi^+\pi^- $ reaction 
\cite{Groom:in}, where the combination of the OZI rule and isospin symmetry
leads 
to an extremely small branching ratio. This can explicitly be seen in 
theoretical calculations 
\cite{Bramon:1979ku,Achasov:1989kz,Genz:zp,Oller:1999ag}, where one finds large 
cancellations as a combined effect of isospin symmetry and the OZI rule. 

   Unlike the $\rho$ coupling to the nucleon which has been the subject of much
research from different theoretical points of view 
\cite{frazer,sakurai,Hohler:1974ht,Brown:1986gu,Ren:1990fx,
Wen:1997rf,Zhu:1999yg,weber,Ferchlander:1982cm,Oset:1983hh}, the $\phi$ 
coupling to the nucleon has comparatively received much less attention. Some
studies done using theoretical dispersion relations give a rather large 
coupling  of the
$\phi$ to the nucleon \cite{hoehler,Hammer:1995de,Hammer:1999uf,Hammer:1998rz} implying a large violation 
of the OZI rule. A reanalysis of the situation was done in 
\cite{Meissner:1997qt}, where the  consideration in the
dispersion-theoretical analysis of the correlated $\rho \pi$ exchange term in
the NN potential \cite{Holinde:1995ux} drastically reduced the former results for the
$\phi$ coupling. At the same time a perturbative calculation  by
explicitly evaluating the indirect coupling of the $\phi$ to the nucleon through
the $K$ and $K^*$ meson cloud and hyperon excitation was done, and it was
concluded that the couplings, although with uncertainties, were indeed small 
and compatible with the expected OZI rule violation.

    New developments in chiral theory and vector meson interaction with nucleons
and nuclei have  given us more elements to tackle the problem and make
a more quantitative evaluation of the $\phi$ coupling. One of the interesting
developments was the combination of chiral symmetry with vector meson dominance
formulated within a gauge invariant framework.  Thanks to this, vertex
corrections of the type of contact terms VPBB
(vector-pseudoscalar-baryon-baryon) are generated 
\cite{Klingl:1997kf,Herrmann:1993za,Urban:1998eg,Urban:2000im,
Cabrera:2000dx} which introduce new terms in
the loop calculations of the vector meson form factors. Such task was undertaken recently in the evaluation of the loop 
contribution to the $\rho$ electric and magnetic form factors \cite{Jido:2002ig}, 
which led to corrections quite stable with respect to moderate changes in the 
regularizing scale of the theory.

   The purpose of the present paper is to
make an evaluation of the electric and magnetic form factors of the $\phi$
coupling to the nucleon for which we follow closely the approach of 
\cite{Jido:2002ig}. There are also other new elements in the present
evaluation, like the consideration of the $\Sigma ^*(1385)$ in addition to the
$\Lambda$ and the $\Sigma$ in the intermediate states. This is done for
consistency with the study of the $\rho$ coupling where $\Delta(1232)$
intermediate states were also considered. The consideration of $\Delta$
intermediate states was advocated in 
\cite{Dashen:1993as,Dashen:1993ac,Jenkins:1993zu} as a way to
implement in the chiral perturbative calculations appropriate limits of large
$N_c$.  The  $\Sigma ^*(1385)$  is the element of the SU(3) decuplet which
plays to the hyperons the role of the $\Delta$ to the nucleons. Although there are other reasons to include the $\Delta$ contribution because of its strong magnetic transition to the nucleon, consistency with $SU(3)$ symmetry suggests the inclusion of the $\Sigma^{*}$ in the strange sector if the $\Delta$ is included in the non strange one.
In fact, as we shall see, the actual calculations show that the contribution
of the $\Sigma^{*}$ is comparable to that of other intermediate hyperons. We also take into account the $\phi-\omega$ mixing which, although with uncertainties, can give a small contribution to the $\phi NN$ vector coupling, but a negligible contribution to the tensor one.

  The strength which we get for the couplings is small and consistent with 
 a weak violation of the OZI rule. On the other hand the results obtained are
 quite stable and provide a realistic determination 
 of the size and sign of the $\phi NN$ electric and magnetic form factors for not
 too large values of the $\phi$ momentum.

\section{Model for the $\phi NN$ coupling}
 
In this section we introduce the Lagrangians needed to calculate the one loop
contributions to the $\phi NN$ couplings and perform the calculation. In general,
the vertex function of the $\phi NN$ coupling can be written in terms of two Lorentz
independent functions, $G^{V}$ and $G^{T}$:

\begin{center}
\be
-it_{\phi NN}=i\left(
G^{V}(q)\gamma^{\mu}+\frac{G^{T}(q)}{2iM_{N}}\sigma^{\mu \nu}q_{\nu}
\right)\epsilon^{*}_{\mu}
\label{rho5}
\ee
\end{center}

\noindent being $q$ and $\epsilon^{*}_{\mu}$ the momentum and the polarization
vector of the outgoing  $\phi$. For convenience, we work in the Breit frame, i.
e., $q^{0}=0$, $\vec{p}_{i}=\vec{q}/2$ (initial proton) and  
$\vec{p}_{f}=-\vec{q}/2$ (final proton), and also in
the non-relativistic limit. Then eq.~(\ref{rho5}) is written as

\begin{center}
\be
-it_{\phi NN}= i
G^{E}(q)\epsilon^{0}-\frac{G^{M}(q)}{2M_{N}}(\vec{\sigma}\times \vec{q})\cdot
\vec{\epsilon}
\label{rho9}
\ee
\end{center}

\noindent with

\begin{center}
\ba
G^{E}(q)& = & G^{V}(q) \nn \\
G^{M}(q)& = & G^{T}(q)+G^{V}(q)
\ea
\end{center}
 
In order to perform the calculations, we use the effective Lagrangians of
refs.~\cite{Klingl:1996by,Klingl:1997kf}, which combine chiral $SU(3)$ dynamics
with VMD \footnote{We have modified the formulae of \cite{Klingl:1996by,Klingl:1997kf}
 in order to use the
normalizations of the $P$, $V_{\mu}$ and $u_{\mu}$ matrices of 
ref.~\cite{Pich:1995bw} which are more
commonly used in the literature when using chiral Lagrangians, and the sign of $g$ to agree
with the paper of the $\rho NN$ coupling of \cite{Jido:2002ig}.}. The basic coupling of the pseudoscalar mesons to the baryons is given
by the Lagrangian

\begin{center}
\be
{\cal L}_{BBP}=\frac{F}{2} tr \left( \bar{B}\gamma_{\mu}\gamma_{5}[u^{\mu},B]\right) +
\frac{D}{2} tr\left( \bar{B}\gamma_{\mu}\gamma_{5}\{u^{\mu},B\}\right)
\label{lagr77}
\ee
\end{center}

\noindent with

\begin{center}
\be
u^{\mu}=-\frac{\sqrt{2}}{f}\left( \partial^{\mu} P
+i\frac{g}{\sqrt{2}}[V^{\mu},P]\right)
\label{lagr78}
\ee
\end{center}

\noindent In these two last equations $B$ and $P$ represent the $SU(3)$ matrix
 fields of the baryon and pseudoscalar meson octets, respectively, 
 $f=93$ MeV is the pion decay constant,  and we take $g=-6.05$, 
$F=0.51$ and $D=0.75$, as done in \cite{Klingl:1997kf}. The vector mesons have been introduced by means of the
minimal substitution scheme, in terms of the matrix $V_{\mu}$, which, when
considering only neutral states, reads

\begin{center}
\be
V_{\mu}= \frac{1}{\sqrt{2}}\textrm{ \ \ }\left( \begin{array}{ccc}
 \rho^{0}_{\mu}+\omega_{\mu} & 0 & 0 \\
 0 & -\rho^{0}_{\mu}+\omega_{\mu} & 0 \\
 0 & 0 & \sqrt{2} \phi_{\mu} \end{array} \right)
\ee
\end{center}

The pseudoscalar meson-vector meson couplings are given in
refs.~\cite{Klingl:1996by,Ecker:1989te}, and can be obtained by introducing a
gauge-covariant derivative in the Klein-Gordon Lagrangian

\begin{center}
\be
\partial_{\mu}P \rightarrow D_{\mu}P=\partial_{\mu}P +
i\frac{g}{\sqrt{2}}[V_{\mu},P]
\label{covariant}
\ee
\end{center}

\noindent In this way we obtain the Lagrangian

\begin{center}
\be
{\cal L}_{VPP}= -\frac{ig}{\sqrt{2}}tr\left(V^{\mu}[
\partial_{\mu}P,P]\right)
\label{lagrVpp}
\ee
\end{center}

The Lagrangian of eq.~(\ref{lagr77}) provides the $BBP$ and $BBVP$
vertices but does not provide the direct couplings of the vector mesons to the baryon fields
$BBV$. These vertices are given by the Lagrangian \cite{Klingl:1997kf}

\begin{center}
\be
{\cal L}_{BBV} =
-\frac{g}{2\sqrt{2}}\left(tr(\bar{B}\gamma_{\mu}[V^{\mu},B]+tr(\bar{B}\gamma_{\mu}B)tr(V^{\mu})
\right)
\label{lagr82}
\ee
\end{center} 

In order to evaluate the contribution of diagrams d) and e) we need the $BBPP$
and $BBPPV$ vertices. These vertices can be obtained from the chiral Lagrangian
\cite{Pich:1995bw}

\begin{center}
\be
{\cal L}= i\textrm{ } tr\left( \bar{B}\gamma^{\mu}[\Gamma_{\mu}, B]\right)
\label{2mesons}
\ee
\end{center}

\noindent with

\begin{center}
\be
\Gamma_{\mu}=\frac{1}{2}\left\{ u^{\dagger}(\partial_{\mu}
+i\frac{g}{2\sqrt{2}}V_{\mu})u + u(\partial_{\mu}
+i\frac{g}{2\sqrt{2}}V_{\mu})u^{\dagger}\right\}
\ee
\end{center}

\noindent where $u$ is defined in reference \cite{Pich:1995bw}.

It is important to stress that, according to the OZI rule, we do not get any 
direct coupling of the $\phi$ to the nucleon
from the Lagrangian in eq.~(\ref{lagr82}). As a consequence, all the 
contributions to the
electric and magnetic form factors of the $\phi$ coupling to the nucleon within this framework should
come from loop diagrams. The one loop diagrams contributing to these form
factors are given in fig.~\ref{loop}. The contribution of each diagram to
$G^{E}_{\phi NN}$ and to $G^{M}_{\phi NN}$ is given in Appendix B. We will discuss
later on with more detail the calculations and results obtained.  
\begin{figure}[ht]
\centering
\epsfysize=6.2cm
\epsfbox{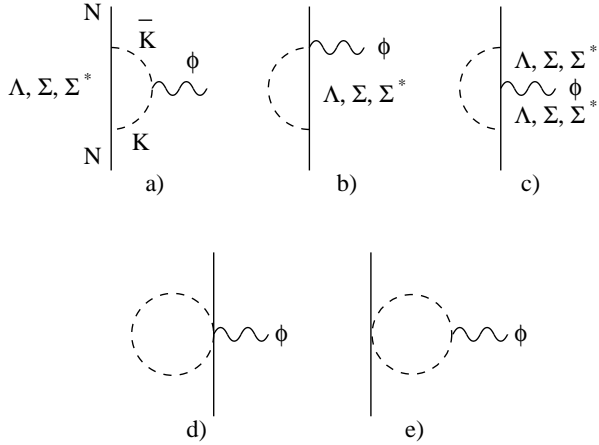}
\caption{One loop diagrams evaluated.}
  \label{loop}
\end{figure}

In the former Lagrangians only baryons from the octet are involved. Here we will
consider also the $\Sigma^{*}$ as an intermediate state, which belongs to the
decuplet. It is worth including this hyperon in our calculations since its
contribution to some processes can be as big as (or even bigger than) the one of
 the $\Sigma$, as can be
seen in reference \cite{Jido:2001am}. The $\bar{K} N\Sigma^{*}$ vertex is given by
\cite{Oset:2000eg}

\begin{center}
\be
V_{\bar{K}N\Sigma^{*}}=\frac{2\sqrt{6}}{5}\frac{D+F}{2f} A\vec{S}^{\dagger} 
\cdot\vec{k}
\label{vertex1}
\ee
\end{center}

\noindent where $\vec{S}^{\dagger}$ is the spin transition operator from $S=1/2$
 to $S=3/2$ and $\vec{k}$ is the momentum of the incoming kaon. The $A$ 
coefficient takes the
values $-1/\sqrt{2}$, $-1$, $-1$, $1/\sqrt{2}$ for the $K^{-}p\rightarrow
\Sigma^{*0}$, $K^{-}n\rightarrow \Sigma^{*-}$, $\bar{K}^{0}
p\rightarrow \Sigma^{*+}$, $\bar{K}^{0}n\rightarrow \Sigma^{*0}$
transitions respectively. To evaluate the diagram b) of fig.~\ref{loop} with
intermediate $\Sigma^{*}$, we need also to know the $\phi N\bar{K}\Sigma^{*}$
 vertices.
These vertices are given in ref. \cite{Oset:2000eg} and have the form

\begin{center}
\be
V_{\phi \bar{K}N\Sigma^{*}} = -g \frac{2\sqrt{3}}{5}\frac{D+F}{2f}
A\vec{S}^{\dagger}\cdot \vec{\epsilon}(\phi)
\label{vertex2}
\ee
\end{center}

\noindent where the $A$ coefficients have the same values as in the 
$\bar{K} N\Sigma^{*}$ vertex. Eqs. (\ref{vertex1}), (\ref{vertex2}) are derived using $SU(6)$ symmetry in \cite{Oset:2000eg}. Since we are in the strange sector, $SU(3)$ could be broken and the most likely way to account for it is through the change $f$ to $f_{K}=1.22f$. We shall also evaluate the results using this latter coupling to estimate uncertainties in the results.

Finally, we will need to know the direct $\phi \Sigma^{*} \Sigma^{*}$ vertex in
order to evaluate diagram c). We can
relate this vertex to the  $\phi \Sigma \Sigma$ one by means of a quark model 
(see Appendix A). 

With the Lagrangians and vertices previously introduced we can evaluate all the
 diagrams in
figure \ref{loop}. In diagram a) we can have $\Lambda$, $\Sigma$ and $\Sigma^{*}$
baryons as intermediate states. The
evaluation of these diagrams is straightforward and the results obtained for
$G^{E}_{\phi NN}(\vec{q}=\vec{0})$ and $G^{M}_{\phi NN}(\vec{q}=\vec{0})$ are shown
in table~\ref{table1} and table~\ref{table2}. 

The other diagrams to be considered contain direct couplings of the $\phi$ to 
the baryonic leg. In these diagrams we multiply the expressions for 
$G^{E}_{\phi NN}(\vec{q})$ and $G^{M}_{\phi NN}(\vec{q})$ given in Appendix B by
 the
$F_{\phi}(\vec{Q})$ form factor, defined in eq.~(\ref{defi}). The contribution 
of
diagram b) to $G^{E}_{\phi NN}(\vec{q})$ is of order ${\cal O}(1/M)$ and we will not
consider it here, as also done in \cite{Jido:2002ig}, since corrections of order ${\cal
O}(1/M)$ in other terms have also been neglected. In Appendix B we give the contributions of diagram b) with
$\Lambda$, $\Sigma$ and $\Sigma^{*}$ as intermediate mesons to $G^{M}_{\phi
NN}(\vec{q})$. The expressions in the
Appendix include the sum of both diagrams b) with the $\phi$ attached to the upper 
and lower vertices.

Another set of diagrams that contribute to both $G^{E}_{\phi NN}(\vec{q})$ and
$G^{M}_{\phi NN}(\vec{q})$ is represented by diagram c) in figure~\ref{loop}, where
$\phi \Lambda \Lambda$, $\phi \Sigma \Sigma$ and $\phi \Sigma^{*} \Sigma^{*}$
vertices appear (we do not have vertices attaching a $\phi$ to two different
baryons, in contrast with the $\rho$ case, since the $\phi$ is an isoscalar). The $\phi \Sigma \Sigma^{*}$ coupling is also zero using $SU(6)$ arguments as done in Appendix A. 
The $\phi \Lambda \Lambda$ and $\phi \Sigma \Sigma$ vertices can be
obtained from Lagrangian~(\ref{lagr82}). However, this Lagrangian does not account
for baryons belonging to the decouplet, and we have to resort to a quark model to
relate the $\phi \Sigma^{*} \Sigma^{*}$ coupling to the $\phi \Sigma \Sigma$, as
announced before. This
is done in Appendix A in an analogous way as it was done in
ref.~\cite{Jido:2002ig} to relate the $\rho
N\Delta$ and $\rho \Delta \Delta$ couplings to the $\rho NN$ coupling. We should also note here that the use of the nonrelativistic $SU(6)$ symmetry to relate these meson baryon baryon couplings was advocated in \cite{Dashen:1993as} in order to ensure basic large $N_c$ counting rules.
 Note also that diagrams a), b) and c), in the case of
intermediate $\Sigma$'s,  account actually for two diagrams, since the 
intermediate
hyperon can be either a $\Sigma^{+}$ or a $\Sigma^{0}$. The same happens in the
case of intermediate $\Sigma^{*}$. 
Finally, diagrams d) and e) of fig.~\ref{loop} do not contribute to the $\phi NN$
coupling at $q=0$ and we do not consider them (see
Appendix B) since we are mostly concerned about the values of the couplings at $q=0$, and the qualitative trend at small $q$ values. We shall further comment on uncertainties from this source. Furthermore, one can also see from the Appendix that these diagrams do not contribute to $G^{T}$. 
  

  \begin{table}
\begin{center}
\begin{tabular}{lllll}
\hline
\hline
Interm. baryon & a) & b) & c) &  Sum \\
\hline
\hline
$\Lambda$ & -0.49 & --- & 0.49 & 0 \\
$\Sigma$ & -0.05 & --- & 0.05 & 0 \\
$\Sigma^{*}$ & 0.40 & --- & -0.40 & 0 \\
\hline
Total $G^{E}_{\phi NN}(\vec{q}=\vec{0})$& & & & 0 \\
\end{tabular}
\caption{Different contributions to $G^{E}_{\phi NN}$ at $\vec{q}=\vec{0}$.}
\label{table1}
\end{center}
\end{table}

  \begin{table}
\begin{center}
\begin{tabular}{lllll}
\hline
\hline
Interm. baryon & a) & b) & c) &  Sum \\
\hline
\hline
$\Lambda$ & -0.75 & 0.60 & -0.17 & -0.32 \\
$\Sigma$ & -0.07 & 0.06 & -0.02 & -0.03 \\
$\Sigma^{*}$ & 0.31 & -0.21 & 1.35 & 1.45 \\
\hline
Total $G^{M}_{\phi NN}(\vec{q}=\vec{0})$& & & & 1.10 \\
\end{tabular}
\caption{Different contributions to $G^{M}_{\phi NN}$ at $\vec{q}=\vec{0}$.}
\label{table2}
\end{center}
\end{table}

It is worth pointing out that the total contribution to $G^{E}_{\phi}(\vec{q})$ at
$\vec{q}=\vec{0}$ is null, due to the cancellation between the contributions of
diagrams a) and c) of fig.~\ref{loop} for each intermediate baryon (we have done the
calculations using an averaged kaon mass $m_{K}= 495.7$ MeV). This cancellation is a
consequence of the gauge symmetry for vector mesons, whose implications were discussed in
detail in~\cite{Jido:2002ig}. These gauge invariance arguments would break in the presence
 of form factors. This is well known in the literature where there are several prescriptions to restore it \cite{Urban:1998eg,Berends:xw,Nacher:1998hh}. The implications of this breaking of gauge invariance due to the presence of form factors were discussed in \cite{Jido:2002ig} in the analogous derivation of the $\rho NN$ coupling. There it was found that $G_{V}^{\rho NN}(q=0)$ was still zero from these mesonic loops even in the presence of form factors. This is also the case here as we can see analytically from the expressions in Appendix B. At $q$ finite there would be a breaking of gauge invariance, but the study of \cite{Jido:2002ig} served to show that the scale at which it is broken is given by the $\Lambda$ parameter of the monopole form factors used (of the order of 1 GeV) and the results for values of $q$ up to about 500 MeV/c were not affected by that symmetry breaking. These results can be extrapolated to the present case, thus limiting the values of $q$ to about the same range.
It is interesting to note that the cancellation of $G^{E}$ at
$q=0$ for the case of the $\rho$ required a term with nucleon wave function
renormalization. This term is null here since the $\phi$ does not couple directly to the
nucleon, but in spite of that, the requirement $G^{E}_{NN\phi}(\vec{q}=0)=0$ also holds here
and comes from a direct cancellation of the terms associated to diagrams a) and c).
  We can also see that the
contributions of the diagrams with an intermediate $\Sigma$ are small compared to
those of the diagrams with intermediate $\Lambda$ or $\Sigma^{*}$. This is 
due to the fact that the contributions of the diagrams with intermediate $\Sigma$
are proportional to $(D-F)^{2}=0.058$, compared to the factors $(D+3F)^{2}=5.20$
 and $(D+F)^{2}=1.59$ in the intermediate $\Lambda$ and $\Sigma^{*}$ cases,
respectively (see Appendix B). Another interesting fact is  
that $G^{M}_{\phi NN}$ is dominated by the contribution of the diagrams with
intermediate $\Sigma^{*}$, specially diagram c) of fig.~\ref{loop} with two
intermediate $\Sigma^{*}$, as we can see in table~\ref{table2}. The value that
we get for this coupling is rather large but still a factor 20 smaller than 
the corresponding factor in the $\rho$ coupling to the nucleon, 
$G^{M,exp}_{\rho NN}\sim 21$. It is also about a factor 6 smaller than the contribution
from loops to $G^{M}_{\rho NN}$ found in~\cite{Jido:2002ig}.
\begin{figure}[ht]
\centering
\epsfysize=12.5cm
\epsfbox{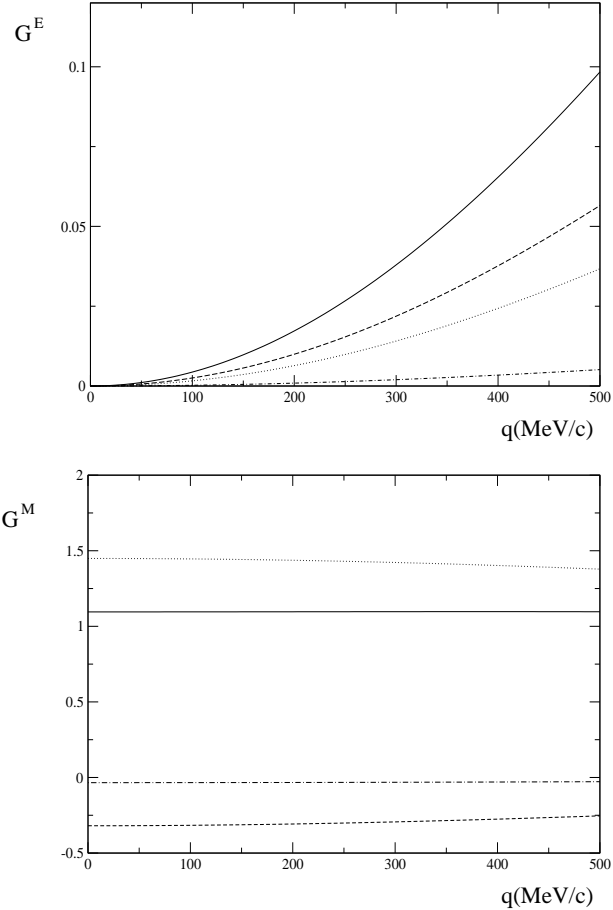}
\caption{q dependence of $G^{E}_{\phi NN}$ and $G^{M}_{\phi NN}$. Dashed line:
diagrams with intermediate $\Lambda$; dashed-dotted line: diagrams with
intermediate $\Sigma$; dotted line: diagrams with intermediate $\Sigma^{*}$;
solid line: sum of all.}
  \label{plots}
\end{figure}

In fig.~\ref{plots} we show the $q$ dependence of both couplings. We only
present our
results up to 500 MeV since we work in the non-relativistic limit and also we
have not taken into account the ${\cal O}(1/M)$ corrections. Furthermore, as discussed above, the presence of form factors would break gauge invariance at the scale of 1 GeV. These approximations
restrict the range of validity of our approach, being the higher
energies region out of the scope of this paper. In the figure we see that
$G^{E}_{\phi NN}$ is null at $\vec{q}=\vec{0}$ and  keeps to small values in the low 
momentum region which we study. Similarly, the $q$ dependence of $G^{M}_{\phi NN}$ 
is very
smooth. Finally, let us stress that the final results have a non negligible
dependence on the input of the calculation, mainly on the value of the
$\Lambda$ parameter in the form factors (see Appendix B). A change of $20\%$ in
this parameter induces a change of the same magnitude in the couplings. This gives an idea of the accuracy of the results.

The work done here does not follow the heavy baryon formalism. This formalism is useful if one wishes to stick to a strict power counting. Hence our evaluation of the loop contribution to the $\phi$ coupling diverts from this strict power counting. On the other hand, at the one loop level of the present calculation it keeps kinetic energies in the propagators, which is also a desirable feature. For some processes, like the meson baryon scattering, where such corrections matter in order to respect thresholds, phase space, which are of relevance to unitarity, etc. \cite{Oset:1997it}, this diversion from the heavy baryon formalism has proved to be phenomenologically advantageous. Although here these effects are no so relevant we have followed the same philosophy, using baryon propagators in their nonrelativistic approximation.

One can also think about including higher order terms which have proved relevant in the unitarization of chiral perturbation theory \cite{Oller:2000ma}. These terms would lead to the $\phi$ coupling to $K\bar{K}$ components and back to the $\phi$, plus iterations of such diagrams. This part has been done in the study of the pion and kaon vector form factors in \cite{Oller:2000ug} and would lead to a dressed $\phi$ propagator in any process where the $\phi$ is exchanged, for instance, between two nucleons, or a nucleon and a kaon. The use of the full $\phi$ propagator as it is obtained in \cite{Oller:2000ug} would be the complement to the work done here with the $\phi NN$ coupling.

\subsection{$\phi - \omega$ mixing contribution.}

\begin{figure}[ht]
\centering
\epsfysize=3.5cm
\epsfbox{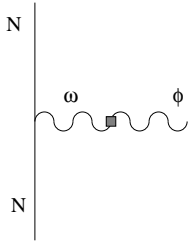}
\caption{Diagram corresponding to the $\phi -\omega$ mixing contribution to the $\phi NN$ coupling.}
  \label{omephimix}
\end{figure}

There is still another contribution to the couplings which we want to address here. This comes from the $\phi - \omega$ mixing and the coupling of the $\omega$ to the nucleon. The $\phi - \omega$ mixing problem has received much attention in the literature and it is still an unsettled problem \cite{Achasov:1989mh,Achasov:1999tp,Achasov:av,Benayoun:2000ti,Nasriddinov:qi,Benayoun:2001qz,LopezCastro:1996xh}. The mechanism producing this new induced $\phi$ coupling to the nucleon is shown in figure~\ref{omephimix}.

The $\omega NN$ coupling has the same structure as the $\phi NN$
 one, eq.~(\ref{rho5}), and it is commonly accepted from studies of the $NN$ interaction that $G_{\omega NN}^{V}$ is of the order of 15 while $G_{\omega NN}^{T}$, unlike the case of the $\rho$,is very small, compatible with zero \cite{Machleidt:hj}. A reanalysis of the isoscalar $NN$ interaction including "$\sigma$" exchange from correlated two pion exchange using a chiral formalism \cite{Oset:2000gn}, plus the uncorrelated two pion exchange of the box diagrams with intermediate $\Delta$, plus $\omega$ exchange, gives a coupling $G_{\omega NN}^{V}=13$ \cite{Jido:2001am}.
 
The $\phi -\omega$ coupling is given by the Lagrangian

\be
{\cal L}=i\Theta_{\phi \omega} \epsilon_{\mu}(\phi)\epsilon^{\mu}(\omega)
\label{mixinglag}
\ee

\noindent where $\Theta_{\phi \omega}$ is the $ \phi -\omega$ mixing parameter and $\epsilon_{\mu}(\phi)$ and $\epsilon_{\mu}(\omega)$ are the polarizations of the $\phi$ and $\omega$ resonances, respectively.

Around the $\omega$, $\phi$ mass there are two possible scenarios \cite{Achasov:1989mh}, one of which has Re $\Theta_{\phi \omega}=0$ (weak mixing) and the other Re $\Theta_{\phi \omega}=20000 \sim 29000$ MeV$^{2}$ (strong mixing). A straightforward evaluation of the diagram of fig.~\ref{omephimix} would provide

\be
G^{V}_{\phi NN}(q=0)=-\frac{\Theta_{\phi \omega}}{M_{\omega}^{2}} G^{V}_{\omega NN}(q=0)\sim -0.6
\label{gmix}
\ee

\noindent in the strong mixing case. This is of the order of $1/5$ of $G^{V}_{\rho NN}$, so still within values compatible with the OZI violation. Yet, this value quoted above, apart from the obvious uncertainties in the choice of possible scenarios, could be irrelevant if one assumes that the $\phi -\omega$ mixing is largely given by the kaon loops as assumed in \cite{Benayoun:2000ti,Benayoun:2001qz}. Should this be the case, then the loop function vanishes at $s=0$ ($q^{0}=0$, $\vec{q}=\vec{0}$ in the present case), as shown in \cite{Benayoun:2000ti} in order to ensure that the photon remains massless \cite{Klingl:1996by}, and also to have current conservation according to \cite{Pichowsky:1999mu}, although strictly speaking it should be sufficient that the sum of loops for different hadrons vanishes at $s=0$. Other arguments in order to support the vanishing at $s=0$ of the individual loops are given \cite{Benayoun:2000ti}. This means that at $q^{0}=q=0$ there would be no contribution to $G^{V}_{\phi NN}(q=0)$ from $\phi -\omega$ mixing. At $q\neq 0$ there could be some contributions, but in the range of momenta considered here the contribution to $G^{V}_{\phi NN}$ using the $\phi -\omega$ mixing given in \cite{Benayoun:2000ti} would be also very small. Of course for $G^{T}_{\phi NN}$, since  $G^{T}_{\omega NN}$ is compatible with zero, there would be no contribution from this mixing.

We have also conducted other tests to estimate uncertainties in the results. First we change $f$ to $f_{K}=1.22 f$ and then all the results are multiplied by the factor $(1.22)^{-2}$, hence multiplying the resuts by 0.67, and producing a $33\%$ reduction of the present results.

On the other hand we have also used different values for the $F$ and $D$ parameters. Appart from those used in the text we have redone the calculations with $D=0.85$, $F=0.52$ \cite{Luty:gi}, and $D=0.80$, $F=0.50$ \cite{Jenkins:1990jv}. In the first of these cases we get $G^{M}(q=0)=1.29$ and in the second case $G^{M}(q=0)=1.16$, to be compared with the number $G^{M}(q=0)=1.10$ that was obtained before using $D=0.75$, $F=0.51$. All this gives us an idea of the uncertainties that we can expect. Together with uncertainties of the order of $20-30\%$ from uncertainties in the form factors, all these sources could lead to about $40\%$ total uncertainty in $G^{M}(q=0)$ when summed in quadrature.
 
\section{Conclusions}  
  
We have evaluated the contributions to $G^{E}_{\phi NN}(\vec{q})$ and   
$G^{M}_{\phi NN}(\vec{q})$ for the OZI violating $\phi NN$ coupling. Since there
is no direct coupling of the $\phi$ to the nucleon, all the contributions come
from loop diagrams, although we have also discussed the effect of $\phi -\omega$ mixing. The loops are regularized by means of a form factor,
introducing an effective cut off of the order of 1.2 GeV. In addition we also
restrict the space of intermediate states to the $\Lambda$, $\Sigma$ and
$\Sigma^{*}$. This kind of regularization from two sources has been 
successfully used in a large number of evaluations of chiral bag models 
\cite{Thomas:1982kv}.

We find that $G^{E}_{\phi NN}$ is null at $q=0$ and grows smoothly in the low
energy regime reaching values of around 0.1 at $q=500$ MeV. In the $G^{M}_{\phi
NN}$ coupling case we find a value of 1.1 at $q=0$, with a very smooth
dependence on the momentum. This coupling is dominated by the contribution of
diagrams with intermediate $\Sigma^{*}$. In both $G^{E}_{\phi NN}$ and $G^{M}_{\phi
NN}$ the contribution of diagrams with intermediate $\Sigma$ is very small
compared to those of the diagrams with intermediate $\Lambda$, $\Sigma^{*}$, due
to the smallness of the $\Sigma KN$, $\Sigma KN\phi$ couplings. The values of
the couplings that we have obtained are small compared to
the corresponding couplings in the case of the $\rho NN$ interaction,
$G^{E}_{\rho NN}=2.9\pm 0.3$ and $G^{M}_{\rho NN}=20.9\pm 2.3$, as expected from
the OZI rule. However, the one loop calculation of $G^{M}_{\rho NN}$ of 
ref.~\cite{Jido:2002ig} gives a value $G^{M}_{\rho NN}=6.05$, only a factor 6
bigger than the one obtained here. 

We have also discussed uncertainties in these form factors. We find that although $G^{V}_{\phi NN}(q=0)$ is equal to zero from loops and also from the $\phi -\omega$ mixing, according to \cite{Benayoun:2000ti}, there would be contributions to   $G^{V}_{\phi NN}(q)$ from the $\phi -\omega$ mixing and also from the diagrams d,e of fig.~\ref{loop} at $q\neq 0$. We have not included these contributions here, hence the $q$ dependence of $G^{V}_{\phi NN}(q)$ obtained from the loops should be only taken as indicative of the trend of the results. On the other hand the results obtained for $G^{T}_{\phi NN}(q)$ are in a more solid ground since neither of the aforementioned mechanisms contributes to this form factor. Hence up to the moderate values of $q\sim 500$ MeV/c, where the presence of the monopole form factors does not spoil gauge invariance, the resuts obtained here should be reliable. Given the weak dependence on $q$ found here for $G^{E}_{\phi NN}$ and $G^{M}_{\phi NN}$ in that range of momenta, the value for $G^{M}_{\phi NN}(q=0)$ and its approximate constancy in that range of momenta should be reliable results, within the uncertainties quoted at the end of the former section.

\subsection*{Acknowledgments}
We would like to thank Ulf-G. Meissner for a careful reading of the
manuscript and useful comments. One of us, J. P. wishes to acknowledge support 
from the Ministerio de Educacion.
This work is also partly
supported by DGICYT contract number BFM2000-1326 and E.U. EURIDICE network
contract no. HPRN-CT-2002-00311. 

\vspace{1cm}

\appendix{\noindent \Large{\bf Appendices}}

\section{Quark model for the $\phi\Sigma^{*} \Sigma^{*} $ vertex}

In this Appendix we relate the $\phi \Sigma^{*} \Sigma^{*}$ and $\phi 
\Sigma \Sigma$ vertices through the $SU(6)$ quark model. Let us define the operator
corresponding to the $\phi$ coupling to the $i$-th quark for $G^{M}$

\begin{center}
\be
\hat{g}^{i}_{M}=-G^{M}_{(q)}\frac{(\vec{q}\times\vec{\epsilon}) \cdot
\vec{\sigma}_{i}}{2m_{q}}
\label{gmin}
\ee
\end{center}

\noindent where $m_{q}$ is the quark mass, $\vec{q}$ is the momentum of the
outgoing $\phi$ and $G^{M}_{(q)}$ is the $G^{M}$ factor corresponding to the $\phi$
coupling to the quark $q$. For a $\Sigma$ baryon with spin up the quark model
provides 

\begin{center}
  \begin{equation}
   \langle \Sigma^{+} \uparrow | \sum_{i=1}^3 \hat{g}_M^{(i)} | \Sigma^{+} \uparrow
   \rangle = -G^{M}_{(q)} \frac{(\vec{q} \times \vec{\varepsilon})_3}{
    2 m_q}
\end{equation}
\end{center}

Here we have used that $\langle \Sigma \uparrow | \sigma_{3} | \Sigma \uparrow
\rangle$=1, as can be obtained using the $\Sigma$ wave function in the spin-flavor
space

\begin{center}
\begin{equation}
    | \Sigma^{+} \uparrow \rangle = {1 \over \sqrt{2}}(\phi_{MS} \chi_{MS} + \phi_{MA}
      \chi_{MA}) 
\end{equation}
\end{center}

\noindent with

\begin{center}
\begin{eqnarray}
   \phi_{MS}^{\Sigma^{+}} = {1 \over \sqrt{6}} [ (us+su)u-2uus] &\hspace{0.5cm}&
   \phi_{MA}^{\Sigma^{+}} = {1 \over \sqrt{2}} (us-su)u\\
   \chi_{MS}^{(\uparrow)} = {1 \over \sqrt{6}} [(\uparrow\downarrow +
   \downarrow\uparrow) \uparrow -2 \uparrow\uparrow\downarrow] &\hspace{0.5cm}&
   \chi_{MA}^{(\uparrow)} = {1 \over \sqrt{2}}
   (\uparrow\downarrow-\downarrow\uparrow)\uparrow 
\end{eqnarray}
\end{center}

Using these wave functions and comparing with the definition of magnetic
coupling to the nucleon (see eq.~(\ref{rho9})), we easily find that

\begin{center}
\be
\frac{G^{M}_{(q)}}{2m_{q}}=\frac{G^{M}_{\Sigma}}{2M_{N}}=-\frac{g}{\sqrt{2}}\frac{1}{2M_{N}}
\label{gmsigqm}
\ee
\end{center}

\noindent where the result $G^{M}_{\Sigma}=-\frac{g}{\sqrt{2}}$ can be obtained
from the Lagrangian of eq.~(\ref{lagr82}).

In the same way, we can relate $G^{M}_{\Sigma^{*}}$ to $G^{M}_{(q)}$. To do that we
must use the $\Sigma^{*+}$ wave function in the spin-flavor space

\begin{center}
\be
| \Sigma^{*+} \uparrow \rangle = | \phi_{S}\rangle | \chi_{S} \rangle
\ee
\end{center}

\noindent with

\begin{center}
\be
   \phi_{S}^{(\Sigma^{*+})} = {1 \over \sqrt{3}} (uus+usu+suu ) \hspace{0.5cm}
   \chi_{S}^{(\uparrow)} = {1 \over \sqrt{3}} (\uparrow\uparrow\downarrow +
   \uparrow\downarrow\uparrow + \downarrow\uparrow\uparrow) 
\ee
\end{center}

Using this wave function we obtain

\begin{center}
\be
   \langle \Sigma^{*+} \uparrow | \sum_{i=1}^3 \hat{g}_M^{(i)} | \Sigma^{*+} \uparrow
   \rangle = -2 G^{M}_{(q)} \frac{(\vec{q} \times \vec{\varepsilon})_3}{
    2 m_q}
\end{equation}
\end{center}

The magnetic coupling to the $\Sigma^{*}$ is defined 

\begin{center}
\be
-it_{\phi\Sigma^{*}\Sigma^{*}} = iG^{E}_{\Sigma^{*}}\epsilon^{0} -
\frac{G^{M}_{\Sigma^{*}}}{2M_{N}} (\vec{S}_{\Sigma^{*}} \times
\vec{q})\textrm{ }\vec{\epsilon}
\ee
\end{center}

Taking care of the normalization of the couplings, it is straightforward to
arrive at

\begin{center}
\be
\frac{G^{M}_{\Sigma^{*}}}{2M_{N}}=4\frac{G^{M}_{(q)}}{2m_{q}}=4\frac{G^{M}_{\Sigma}}{2M_{N}}
=-2\sqrt{2}g \frac{1}{2M_{N}}
\ee
\end{center}

The evaluation of $G^{E}_{\Sigma^{*}}$ is analogous and even easier since in
this case only matrix elements of the identity in both the spin and flavor space
must be calculated. We get

\begin{center}
\be
G^{E}_{\Sigma^{*}}=G^{E}_{\Sigma}=-\frac{g}{\sqrt{2}}
\ee
\end{center}

We do not get any $\phi \Sigma \Sigma^{*}$ model from $SU(6)$ symmetry since in the evaluation we get the scalar products of the flavor wave functions: $\langle \phi_{S}^{\Sigma^{*}} | \phi_{MS}^{\Sigma} \rangle$ and $\langle \phi_{S}^{\Sigma^{*}} | \phi_{MA}^{\Sigma} \rangle$ which are null.

\section{ One loop calculations}

In this Appendix we give the explicit expressions of the contributions of the
loop diagrams to $G^{E}_{NN\phi}$ and $G^{M}_{NN\phi}$. In the following
equations and diagrams 
$\epsilon_{\mu}$ denotes the $\phi$ polarization vector, and:

\begin{center}
\ba
q\equiv (E(\vec{q}),\vec{q}) \textrm{\ \ \ \ \ \ \ } & & Q\equiv (0,\vec{q}) \nn \\
  \omega(k)\equiv\sqrt{\vec{k}^{2}+m_{K}^{2}} \textrm{\ \ \ \ \ }& &
  D(k)\equiv\frac{1}{k^{2}-m_{K}^{2}} \nn \\
F_{K}(\vec{k})\equiv \frac{\Lambda^{2}}{\Lambda^{2}+\vec{k}^{2}} \textrm{ \ \ \ \ }
& &F_{\phi}(\vec{k})\equiv \frac{\Lambda_{\phi}^{2}}
{\Lambda_{\phi}^{2}+\vec{k}^{2}}\nn \\
\Lambda=1.2\textrm{ GeV \ \ \ \ \ \ \ \ }  & &\Lambda_{\phi}=2.5\textrm{ GeV}\nn \\
E(\vec{k})\equiv\frac{\vec{k}^{\textrm{ }2}}{2M_{N}}+M_{N}\textrm{\ \ \ \ \ \ \ \ \ } & & 
E_{Y}(\vec{k})\equiv \frac{\vec{k}^{\textrm{ }2}}{2M_{Y}}+M_{Y}
\label{defi}
\ea
\end{center}

\noindent where the subindex $Y$ refers to any of the hyperons considered here
($\Lambda$, $\Sigma$, $\Sigma^{*}$). The values of the $\Lambda$ parameters appearing in the form factors (1.2 GeV for the coupling of pseudoscalars and 2.5 GeV for the coupling of vector mesons) are motivated by the study of the $NN$ interaction in \cite{Machleidt:hj} and are taken there for pions and $\rho$ mesons by analogy. 
We warn the reader that, in order not to complicate excessively the expressions,
we have deliberately omitted the form factors and the $M/E$ relativistic
corrections to the baryonic propagators in the following equations,
although it should be kept in mind that one must include them to perform the
numerical calculations.


\hspace{-1cm}\parbox{3cm}{
\psfig{file=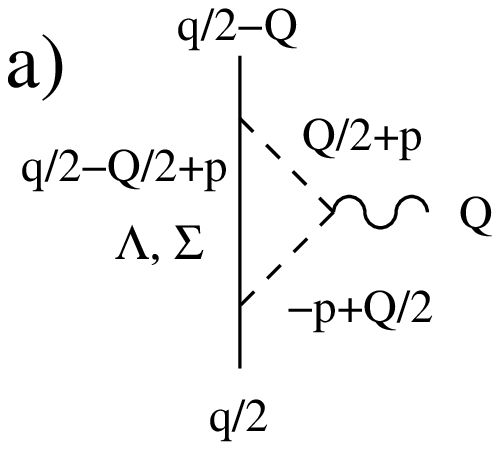,width=0.25\textwidth,silent=}

\label{geNfig}
}
\parbox{10cm}{
\be
G^{E\textrm{ }a)}_{\phi NN}(\vec{q})=\alpha^{a)}_{Y}g 
\int\frac{d^{3}p}{(2\pi)^{3}}\left(\vec{p}^{\textrm{ }2}-\frac{\vec{q}^{\textrm{ }2}}{4}\right)
f_{1}(\vec{p},\vec{q})
\label{geN}
\ee}
\begin{center}
\be
\frac{G^{M \textrm{ }a)}_{\phi NN}}{2M_{N}}=\alpha^{a)}_{Y}g  
\int\frac{d^{3}p}{(2\pi)^{3}}\left(\vec{p}^{\textrm{ }2}-\frac{(\vec{p}
\vec{q})^{2}}{\vec{q}^{\textrm{ }2}}\right) f_{2}(\vec{p},\vec{q})
\label{gmN}
\ee
\end{center}

\noindent where the $\alpha_{Y}$ and $\beta_{Y}$ coefficients, for $\Lambda$
and $\Sigma$ intermediate hyperons, are:

\begin{center}
\be
\alpha^{a)}_{\Lambda}=\frac{1}{3\sqrt{2}}\left(\frac{D+3F}{2f}\right)^{2} \textrm{ \
\ \ \ }; 
  \alpha^{a)}_{\Sigma}= \frac{3}{\sqrt{2}}\left(\frac{D-F}{2f}\right)^{2} 
\label{betaa}
\ee
\end{center}

 The $f_1$ and $f_2$ functions are defined as:

\begin{center}
\ba
f_{1}(\vec{p},\vec{q})&=&\frac{1}{\omega(\vec{p}+\vec{q}/2)+\omega(\vec{p}-
\vec{q}/2)}\textrm{ }\frac{1}{E(\vec{q}/2)-\omega(\vec{p}+\vec{q}/2)-E_{Y}(\vec{p})}
\times \label{efes} \\ & \times&\frac{1}{E(\vec{q}/2)-\omega(\vec{p}-\vec{q}/2)-
E_{Y}(\vec{p})}\nn \\
f_{2}(\vec{p},\vec{q})&=&\frac{1}{\omega(\vec{p}+\vec{q}/2)+\omega(\vec{p}-
\vec{q}/2)}\textrm{ }\frac{1}{E(\vec{q}/2)-\omega(\vec{p}+\vec{q}/2)-E_{Y}(\vec{p})}
\times \nn \\ &\times &\frac{\omega
(\vec{p}+\vec{q}/2)+\omega(\vec{p}-\vec{q}/2)+E_{Y}(\vec{p})-E(\vec{q}/2)}
{E(\vec{q}/2)-\omega(\vec{p}-\vec{q}/2)-E_{Y}(\vec{p})}
 \textrm{ }\frac{1}{2\omega(\vec{p}+\vec{q}/2)\omega(\vec{p}-\vec{q}/2)}
\nn
\ea
\end{center}
\vspace{0.3cm}

\begin{figure}[h]
\psfig{file=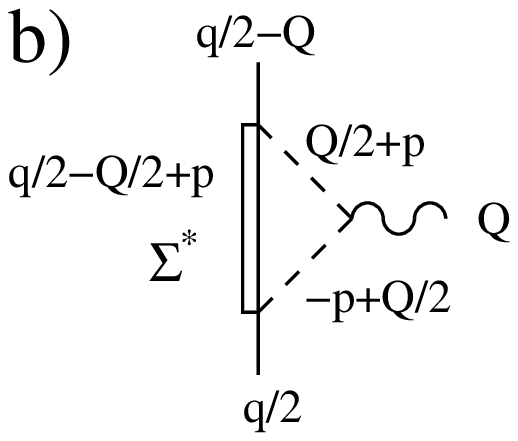,width=0.25\textwidth,silent=}
\label{gedelfig}
\end{figure} 

In the calculation of diagrams with intermediate $\Sigma^{*}$'s one has different spin
and isospin factors since the spin and isospin transition operators appearing in
the corresponding Lagrangians satisfy the following relations:

\begin{center}
\begin{eqnarray}
S_{i}S^{\dagger}_{j}=\frac{2}{3}\delta_{ij}-
\frac{i}{3}\epsilon_{ijk}\sigma_{k}\nn \\
\label{repchange}
\ea
\end{center}

Taking this into account one finds

\begin{center}
\be
G^{E\textrm{ }b)}_{\phi NN}(\vec{q})=\frac{12\sqrt{2}}{25} \left(\frac{D+F}{2f}\right)^{2} g
\int\frac{d^{3}p}{(2\pi)^{3}}\left(\vec{p}^{\textrm{ }2}-\frac{\vec{q}^{\textrm{
}2}}{4}\right)
f_{1}(\vec{p},\vec{q})
\label{gedel}
\ee
\end{center}

\begin{center}
\be
\frac{G^{M\textrm{ }b)}_{\phi NN}}{2M_{N}}=-\frac{6\sqrt{2}}{25} \left(\frac{D+F}{2f}\right)^{2} g 
\int\frac{d^{3}p}{(2\pi)^{3}}\left(\vec{p}^{\textrm{ }2}-\frac{(\vec{p}
\vec{q})^{2}}{\vec{q}^{\textrm{ }2}}\right) f_{2}(\vec{p},\vec{q})
\label{gmdel}
\ee
\end{center}

\vspace{0.6cm}

\hspace{-1cm}\parbox{3cm}{
\psfig{file=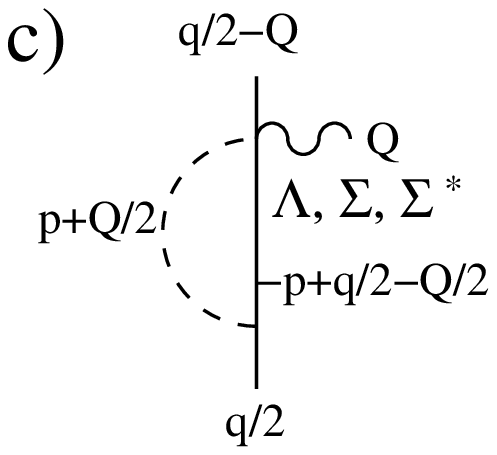,width=0.25\textwidth,silent=}

\label{gepefig}
}
\parbox{10cm}{
\be
G^{E\textrm{ }c)}_{\phi NN}(\vec{q})={\cal O}(1/M_{N})
\ee}

\begin{center}
\ba
\textrm{ \ }\frac{G^{M\textrm{ }c)}_{\phi NN}}{2M_{N}}&=-&\beta^{c)}_{Y} g
\int\frac{d^{3}p}{(2\pi)^{3}}\left(1+\frac{2\vec{p}\vec{q}}{\vec{q}^{2}}\right)\frac{1}{2\omega
(\vec{p}+\vec{q}/2)}\times \nn \\ &\times & \frac{1}{E(\vec{q}/2)-
\omega(\vec{p}+\vec{q}/2)-E_{Y}(\vec{p})}
\label{gmpe}
\ea
\end{center}

\noindent with

\begin{center}
\ba
\beta^{c)}_{\Lambda}=-\frac{1}{3\sqrt{2}}\left(\frac{D+3F}{2f}\right)^{2}&; \textrm{
\ \  \  \ \ \ }
\beta^{c)}_{\Sigma}=-\frac{3}{\sqrt{2}}\left(\frac{D-F}{2f}\right)^{2} ;  &\nn \\
\beta^{c)}_{\Sigma^{*}}=\frac{6\sqrt{2}}{25}\left(\frac{D+F}{2f}\right)^{2} 
\label{betac}
\ea
\end{center}
\vspace{0.6cm}

\hspace{-1cm}\parbox{3cm}{
\psfig{file=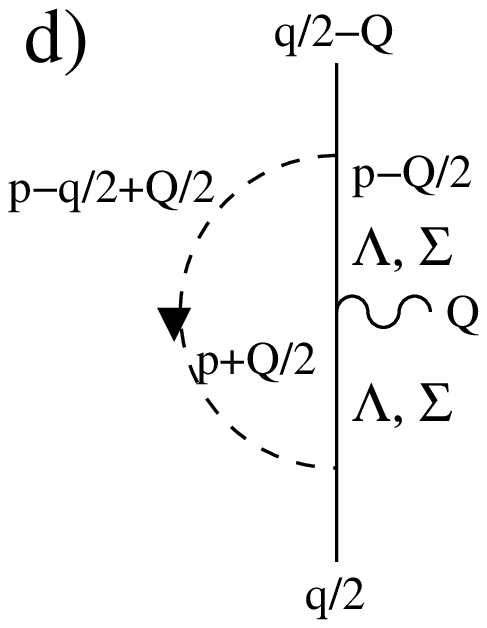,width=0.20\textwidth,silent=}
\label{ge1fig}
}
\parbox{10cm}{
\ba
\textrm{ \ \ }G^{E\textrm{ }d)}_{\phi NN}(\vec{q})=-\alpha^{d)}_{Y}g
\int\frac{d^{3}p}{(2\pi)^{3}}\frac{1}{2\omega(\vec{p})} 
\times \nn \\ \times \frac{\vec{p}^{\textrm{
}2}}{E(\vec{q}/2)-\omega(\vec{p})-E_{Y}(\vec{q}/2-\vec{p})}
\frac{1}{ E(\vec{q}/2)-\omega(\vec{p})-E_{Y}(-\vec{q}/2-\vec{p})}
\label{ge1}
\ea
}
\begin{center}
\ba
G^{M\textrm{ }d)}_{\phi NN}(\vec{q})&=&\alpha^{d)}_{Y} g \int\frac{d^{3}p}
{(2\pi)^{3}}\frac{(\vec{p}\vec{q})^{2}}{2\vec{q}^{\textrm{ }2}\omega(\vec{p})}
 \frac{1}{E(\vec{q}/2)-\omega(\vec{p})-E_{Y}(\vec{q}/2-\vec{p})}
\times \nn \\ &\times&\frac{1}{ E(\vec{q}/2)-\omega(\vec{p})-E_{Y}(-\vec{q}/2-\vec{p})}
\label{gm1}
\ea
\end{center}

\begin{center}
\be
\alpha_{\Lambda}^{d)}=\frac{1}{3\sqrt{2}}\left(\frac{D+3F}{2f}\right)^{2} \textrm{ \
\ \ \ }; 
  \alpha_{\Sigma}^{d)}= \frac{3}{\sqrt{2}}\left(\frac{D-F}{2f}\right)^{2}  
\label{betad}
\ee
\end{center}

\hspace{-1cm}\parbox{3cm}{
\psfig{file=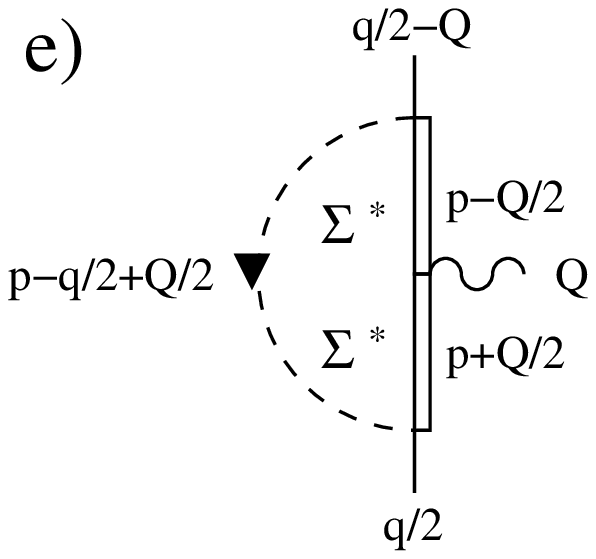,width=0.25\textwidth,silent=}
\label{ge1dfig}
}
\parbox{10cm}{\ba
\textrm{\hspace{-0.3cm}}G^{E\textrm{ }e)}_{\phi
NN}(\vec{q})=-\frac{12\sqrt{2}}{25}\left(\frac{D+F}{2f}\right)^{2}g
\int\frac{d^{3}p}{(2\pi)^{3}}\frac{1}{2\omega(\vec{p})} \times 
 \label{ge1d} \\ \times 
\frac{\vec{p}^{\textrm{ }2}}{E(\vec{q}/2)-\omega(\vec{p})-E_{\Sigma^{*}}(\vec{q}/2-
\vec{p})}\textrm{ }
\frac{1}{ E(\vec{q}/2)-\omega(\vec{p})-E_{\Sigma^{*}}(-\vec{q}/2-\vec{p})}
\nn
\ea}

\begin{center}
\ba
G^{M\textrm{ }e)}_{\phi
NN}(\vec{q})&=&-\frac{36\sqrt{2}}{25}g\left(\frac{D+F}{2f}\right)^{2} 
\int\frac{d^{3}p}{(2\pi)^{3}}
\left( \vec{p}^{\textrm{ }2}+\frac{(\vec{p}\vec{q})^{2}}{3\vec{q}^{\textrm{ }2}}
\right) \frac{1}{2\omega(\vec{p})} \times  \nn \\
&\times& \frac{1}{E(\vec{q}/2)-\omega(\vec{p})-E_{\Sigma^{*}}(\vec{q}/2-\vec{p})}\textrm{ }
\frac{1}{E(\vec{q}/2)-\omega(\vec{p})-E_{\Sigma^{*}}(-\vec{q}/2-\vec{p})}
\nn \\ 
\label{gm1d}
\ea
\end{center}

To evaluate this diagram we have used the relation

\begin{center}
\be
S_{i}S_{j}S^{\dagger}_{k}=\frac{5}{6}\textrm{i}\epsilon_{ijk}-
\frac{1}{6}\delta_{ij}\sigma_{k}+\frac{2}{3}\delta_{ik}\sigma_{j}- \frac{1}{6}
\delta_{jk}\sigma_{i}
\label{releses}
\ee
\end{center}

\vspace{0.3cm}

\vspace{0.3cm}
\begin{figure}[h]
\psfig{file=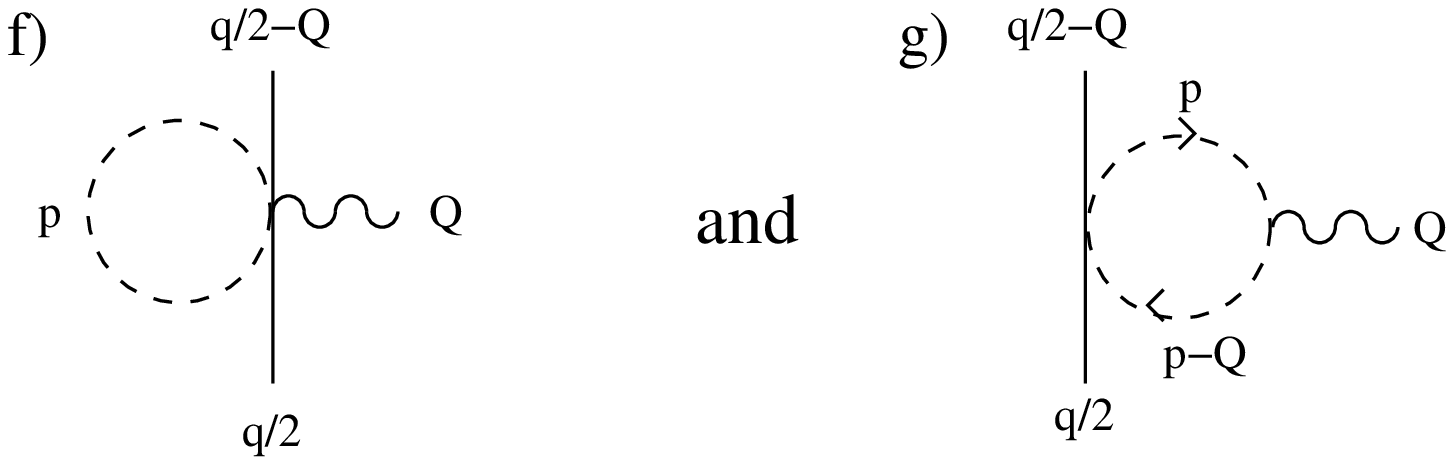,width=0.68\textwidth,silent=}
\label{cance}
\end{figure} 

We do not take into account diagrams f) and g) since they cancel at
$\vec{q}=\vec{0}$. At this value of $\vec{q}$ diagram f) is proportional to:

\begin{center}
\begin{equation}
\int\frac{d^{4}p}{(2\pi)^{4}}2\gamma^{\mu}g_{\mu\nu}\epsilon^{\nu }D(p)
\end{equation}
\end{center}

\noindent and diagram g) is proportional to:

\begin{center}
\begin{equation}
-\int\frac{d^{4}p}{(2\pi)^{4}}\gamma^{\mu}4p_{\mu}\epsilon^{\nu }p_{\nu}D(p)D(p)
\label{jyk}
\end{equation}
\end{center}

Taking into account the integral identity:

\begin{center}
\be
\int {d^{4}p}\frac{4p^{\mu}p^{\nu}}{(p^{2}+s+i\epsilon)^{2}} = \int
d^{4}p\frac{2g^{\mu\nu}}{k^{2}+s+i\epsilon}
\label{relation}
\ee
\end{center}

\noindent it is straightforward to see that these diagrams cancel at
$\vec{q}=\vec{0}$.


\begin{thebibliography}{99}




\bibitem{Kaymakcalan:1984bz}
O.~Kaymakcalan and J.~Schechter,
Phys.\ Rev.\ D {\bf 31} (1985) 1109.

\bibitem{Borasoy:1995ds}
B.~Borasoy and U.-G.~Meissner,
Int.\ J.\ Mod.\ Phys.\ A {\bf 11} (1996) 5183
[arXiv:hep-ph/9511320].

\bibitem{Klingl:1996by}
F.~Klingl, N.~Kaiser and W.~Weise,
Z.\ Phys.\ A {\bf 356} (1996) 193
[arXiv:hep-ph/9607431].

\bibitem{Klingl:1997kf}
F.~Klingl, N.~Kaiser and W.~Weise,
Nucl.\ Phys.\ A {\bf 624} (1997) 527
[arXiv:hep-ph/9704398].

\bibitem{Groom:in}
D.~E.~Groom {\it et al.}  [Particle Data Group Collaboration],
Eur.\ Phys.\ J.\ C {\bf 15} (2000) 1.

\bibitem{Bramon:1979ku}
A.~Bramon and A.~Varias,
Phys.\ Rev.\ D {\bf 20} (1979) 2262.

\bibitem{Achasov:1989kz}
N.~N.~Achasov and A.~A.~Kozhevnikov,
Phys.\ Lett.\ B {\bf 233} (1989) 474.

\bibitem{Genz:zp}
H.~Genz and S.~Tatur,
Phys.\ Rev.\ D {\bf 50} (1994) 3263
[arXiv:hep-ph/9401263].

\bibitem{Oller:1999ag}
J.~A.~Oller, E.~Oset and J.~R.~Pelaez,
Phys.\ Rev.\ D {\bf 62} (2000) 114017
[arXiv:hep-ph/9911297].

\bibitem{frazer}
W. R. Frazer and J. R. Fulco, Phys. Rev. {\bf 117} (1960) 1603; {\bf 117} (1960) 1609

\bibitem{sakurai}
J. J. Sakurai, \emph{Currents and Mesons} (University of Chicago Press, Chicago, 1969)


\bibitem{Hohler:1974ht}
G.~H\"ohler and E.~Pietarinen,
Nucl.\ Phys.\ B {\bf 95} (1975) 210.



\bibitem{Brown:1986gu}
G.~E.~Brown, M.~Rho and W.~Weise,
Nucl.\ Phys.\ A {\bf 454} (1986) 669.

\bibitem{Ren:1990fx}
C.~Y.~Ren and M.~K.~Banerjee,
Phys.\ Rev.\ C {\bf 41} (1990) 2370.

\bibitem{Wen:1997rf}
C.~Y.~Wen and W.~Y.~Hwang,
Phys.\ Rev.\ C {\bf 56} (1997) 3346.

\bibitem{Zhu:1999yg}
S.~L.~Zhu,
Phys.\ Rev.\ C {\bf 59} (1999) 435
[arXiv:nucl-th/9809032].

\bibitem{weber} B.L.G. Bakker, M. Bozoian, J.N. Maslow and H.J. Weber, Phys.
Rev. C {\bf 25} (1982) 1134; H.J. Weber, Phys. Lett. B {\bf 233} (1989) 267.

\bibitem{Ferchlander:1982cm}
W.~Ferchlander,
Phys.\ Rev.\ D {\bf 25} (1982) 1432.

\bibitem{Oset:1983hh}
E.~Oset,
Nucl.\ Phys.\ A {\bf 430} (1984) 713.

\bibitem{hoehler} G. H\"ohler et al., Nucl. Phys. B {\bf 114} (1976) 505

\bibitem{Hammer:1995de}
H.~W.~Hammer, U.-G.~Meissner and D.~Drechsel,
Phys.\ Lett.\ B {\bf 367} (1996) 323
[arXiv:hep-ph/9509393].
P.~Mergell, U.-G.~Meissner and D.~Drechsel,
Nucl.\ Phys.\ A {\bf 596} (1996) 367
[arXiv:hep-ph/9506375].

\bibitem{Hammer:1999uf}
H.~W.~Hammer and M.~J.~Ramsey-Musolf,
Phys.\ Rev.\ C {\bf 60} (1999) 045204
[Erratum-ibid.\ C {\bf 62} (2000) 049902]
[arXiv:hep-ph/9903367].

\bibitem{Hammer:1998rz}
H.~W.~Hammer and M.~J.~Ramsey-Musolf,
Phys.\ Rev.\ C {\bf 60} (1999) 045205
[Erratum-ibid.\ C {\bf 62} (2000) 049903]
[arXiv:hep-ph/9812261].



\bibitem{Meissner:1997qt}
U.-G.~Meissner, V.~Mull, J.~Speth and J.~W.~van Orden,
Phys.\ Lett.\ B {\bf 408} (1997) 381
[arXiv:hep-ph/9701296].

\bibitem{Holinde:1995ux}
K.~Holinde,
Prog.\ Part.\ Nucl.\ Phys.\  {\bf 36} (1996) 311
[arXiv:nucl-th/9512001].

\bibitem{Herrmann:1993za}
M.~Herrmann, B.~L.~Friman and W.~Norenberg,
Nucl.\ Phys.\ A {\bf 560} (1993) 411.

\bibitem{Urban:1998eg}
M.~Urban, M.~Buballa, R.~Rapp and J.~Wambach,
Nucl.\ Phys.\ A {\bf 641} (1998) 433
[arXiv:nucl-th/9806030].

\bibitem{Urban:2000im}
M.~Urban, M.~Buballa and J.~Wambach,
Nucl.\ Phys.\ A {\bf 673} (2000) 357
[arXiv:nucl-th/9910004].

\bibitem{Cabrera:2000dx}
D.~Cabrera, E.~Oset and M.~J.~Vicente Vacas,
Nucl.\ Phys.\ A {\bf 705} (2002) 90
[arXiv:nucl-th/0011037].

\bibitem{Jido:2002ig}
D.~Jido, E.~Oset and J.~E.~Palomar,
Nucl.\ Phys.\ A {\bf 709} (2002) 345
[arXiv:nucl-th/0202070].
\bibitem{Dashen:1993as}
R.~F.~Dashen and A.~V.~Manohar,
Phys.\ Lett.\ B {\bf 315} (1993) 425
[arXiv:hep-ph/9307241].

\bibitem{Dashen:1993ac}
R.~F.~Dashen and A.~V.~Manohar,
Phys.\ Lett.\ B {\bf 315} (1993) 438
[arXiv:hep-ph/9307242].

\bibitem{Jenkins:1993zu}
E.~Jenkins,
Phys.\ Lett.\ B {\bf 315} (1993) 441
[arXiv:hep-ph/9307244].

\bibitem{Ecker:1989te}
G.~Ecker, J.~Gasser, A.~Pich and E.~de Rafael,
Nucl.\ Phys.\ B {\bf 321} (1989) 311.


\bibitem{Pich:1995bw}
A.~Pich,
Rept.\ Prog.\ Phys.\  {\bf 58} (1995) 563
[arXiv:hep-ph/9502366].

\bibitem{Jido:2001am}
D.~Jido, E.~Oset and J.~E.~Palomar,
Nucl.\ Phys.\ A {\bf 694} (2001) 525
[arXiv:nucl-th/0101051].


\bibitem{Oset:2000eg}
E.~Oset and A.~Ramos,
Nucl.\ Phys.\ A {\bf 679} (2001) 616
[arXiv:nucl-th/0005046].






















































\bibitem{Berends:xw}
F.~A.~Berends and R.~Gastmans,
Phys.\ Rev.\ D {\bf 5} (1972) 204.

\bibitem{Nacher:1998hh}
J.~C.~Nacher and E.~Oset,
Nucl.\ Phys.\ A {\bf 674}, 205 (2000)
[arXiv:nucl-th/9804006].

\bibitem{Oset:1997it}
E.~Oset and A.~Ramos,
Nucl.\ Phys.\ A {\bf 635} (1998) 99
[arXiv:nucl-th/9711022].

\bibitem{Oller:2000ma}
J.~A.~Oller, E.~Oset and A.~Ramos,
Prog.\ Part.\ Nucl.\ Phys.\  {\bf 45} (2000) 157
[arXiv:hep-ph/0002193].

\bibitem{Oller:2000ug}
J.~A.~Oller, E.~Oset and J.~E.~Palomar,
Phys.\ Rev.\ D {\bf 63} (2001) 114009
[arXiv:hep-ph/0011096].

\bibitem{Achasov:1989mh}
N.~N.~Achasov, M.~S.~Dubrovin, V.~N.~Ivanchenko, A.~A.~Kozhevnikov and E.~V.~Pakhtusova,
Sov.\ J.\ Nucl.\ Phys.\  {\bf 54} (1991) 664
[Yad.\ Fiz.\  {\bf 54} (1991\ IMPAE,A7,3187-3202.1992) 1097].

\bibitem{Achasov:1999tp}
N.~N.~Achasov and A.~A.~Kozhevnikov,
Phys.\ Rev.\ D {\bf 61} (2000) 054005
[arXiv:hep-ph/9906520].

\bibitem{Achasov:av}
N.~N.~Achasov and A.~A.~Kozhevnikov,
Phys.\ Atom.\ Nucl.\  {\bf 63} (2000) 1936
[Yad.\ Fiz.\  {\bf 63} (2000) 2029].

\bibitem{Benayoun:2000ti}
M.~Benayoun, L.~DelBuono, P.~Leruste and H.~B.~O'Connell,
Eur.\ Phys.\ J.\ C {\bf 17} (2000) 303
[arXiv:nucl-th/0004005].

\bibitem{Nasriddinov:qi}
K.~R.~Nasriddinov, B.~N.~Kuranov, G.~G.~Takhtamyshev and T.~A.~Merkulova,
Phys.\ Atom.\ Nucl.\  {\bf 64} (2001) 1326
[Yad.\ Fiz.\  {\bf 64} (2001) 1402].


\bibitem{Benayoun:2001qz}
M.~Benayoun and H.~B.~O'Connell,
Eur.\ Phys.\ J.\ C {\bf 22} (2001) 503
[arXiv:nucl-th/0107047].

\bibitem{LopezCastro:1996xh}
G.~Lopez Castro and D.~A.~Lopez Falcon,
Phys.\ Rev.\ D {\bf 54} (1996) 4400
[arXiv:hep-ph/9607409].




\bibitem{Machleidt:hj}
R.~Machleidt, K.~Holinde and C.~Elster,
Phys.\ Rept.\  {\bf 149} (1987) 1.

\bibitem{Oset:2000gn}
E.~Oset, H.~Toki, M.~Mizobe and T.~T.~Takahashi,
Prog.\ Theor.\ Phys.\  {\bf 103} (2000) 351
[arXiv:nucl-th/0011008].


\bibitem{Pichowsky:1999mu}
M.~A.~Pichowsky, S.~Walawalkar and S.~Capstick,
Phys.\ Rev.\ D {\bf 60} (1999) 054030
[arXiv:nucl-th/9904079].

\bibitem{Luty:gi}
M.~A.~Luty and M.~J.~White,
Phys.\ Lett.\ B {\bf 319} (1993) 261.

\bibitem{Jenkins:1990jv}
E.~Jenkins and A.~V.~Manohar,
Phys.\ Lett.\ B {\bf 255} (1991) 558.


\bibitem{Thomas:1982kv}
A.~W.~Thomas,
Adv.\ Nucl.\ Phys.\  {\bf 13}, 1 (1984).


\end{thebibliography}
\end{document}